\documentclass{mnras}

\usepackage{graphicx}
\usepackage{subfigure}
\usepackage{txfonts}
\let\bibhang\relax
\usepackage{natbib}
\bibpunct{(}{)}{;}{a}{}{,}

\title[A \textit{Herschel} resolved debris disc around HD 105211]{A \textit{Herschel}\thanks{\textit{Herschel} is an ESA space observatory with science instruments provided by the European-led Principle Investigator and with important participation from NASA}  resolved debris disc around HD 105211}

\author[S. Hengst et. al.]{S. Hengst$^{1}$\thanks{Contact e-mail: \href{mailto:U1082091@umail.usq.edu.au}{shane.hengst@usq.edu.au}}, {J. P. Marshall$^{2,3,1} $}, \ {J. Horner$^{1,3}$},\  and {S. C. Marsden$^{1}$}
\\
$^{1}$Computational Engineering and Science Research Centre, University of Southern Queensland, Toowoomba, QLD 4350, Australia\\
$^{2}$School of Physics, UNSW Australia, High Street, Kensington, NSW 2052, Australia\\
$^{3}$Australian Centre for Astrobiology, UNSW Australia, High Street, Kensington, NSW 2052, Australia}

\date{Last updated ---; in original form ---}

\pubyear{2016}

\begin{document}
\label{firstpage}
\pagerange{\pageref{firstpage}--\pageref{lastpage}}
\maketitle

\begin{abstract}
Debris discs are the dusty aftermath of planet formation processes around main-sequence stars.  Analysis of these discs is often hampered by the absence of any meaningful constraint on the location and spatial extent of the disc around its host star.  Multi-wavelength, resolved imaging ameliorates the degeneracies inherent in the modelling process, making such data indispensable in the interpretation of these systems.  \\ The \textit{Herschel} Space Observatory observed HD 105211 ($\eta$ Cru, HIP~59072) with its PACS instrument in three far-infrared wavebands (70, 100 and 160~$\mu$m).  Here we combine these data with ancillary photometry spanning optical to far-infrared wavelengths in order to determine the extent of the circumstellar disc.  The spectral energy distribution and multi-wavelength resolved emission of the disc are simultaneously modelled using a radiative transfer and imaging codes.  Analysis of the \textit{Herschel}/PACS images reveals the presence of extended structure in all three PACS images. From a radiative transfer model we derive a disc extent of 87.0 $\pm$ 2.5 au, with an inclination of 70.7 $\pm$ $2.2 \degr$ to the line of sight and a position angle of 30.1 $\pm$ $0.5 \degr$.  Deconvolution of the \textit{Herschel} images reveal a potential asymmetry but this remains uncertain as a combined radiative transfer and image analysis replicate both the structure and the emission of the disc using a single axisymmetric annulus.
\end{abstract}

\begin{keywords}
stars: individual: HD 105211 -- circumstellar matter -- infrared: stars: planetary systems -- disc-planet interactions
\newline
\newline
\textbf{Accepted to appear in the Monthly Notices of the Royal Astronomical Society on 25th March, 2017}
\end{keywords}



\section{Introduction}

In the early 1980s, the \textit{InfraRed Astronomical Satellite} \citep[\textit{IRAS}; ][]{1984Neugebauer}, announced the surprising discovery that several young, nearby stars were brighter at infrared wavelengths than was expected \citep[e.g.][]{1984Aumann,1984Harper}.  It was soon realised that this excess infrared emission was the signature of dusty debris in orbit around those stars; being analogous to, but far more massive than the Solar system's Asteroid Belt \citep[e.g.][]{1990Aumann,1993Zucker,2010Nesvorny}.  In the three decades since, the detection of such infrared excesses has become routine; with many main sequence stars being found to host significant amounts of circumstellar debris at mid- and far-infrared wavelengths \citep[e.g.][]{2006Su,2013Fujiwara,2013Eiroa,2013KenWya,2014Thureau,2014Ballering,2016Mont}.

We now know that such debris discs are a normal result of the stellar and planetary formation process.  Observations have shown that most of the youngest, low-mass stars are attended by massive gas- and dust-rich circumstellar discs \citep{2011WilliamsCieza}.  It is within these discs that the formation of planetary systems like our own occurs, assembling planetesimals and planets \citep{1993Lissauer} from dust grains over the first 1 to 100~Myr of the star's life \citep{2002Meyer,2009Carpenter,2011Olive}. These protoplanetary discs undergo significant evolution over the first 10 Myr of the star's life, which can be traced through the decay of excess emission at near- and mid-infrared wavelengths \cite[e.g.][]{2014Ribas}, and a corresponding drop in emission at sub-millimetre wavelengths \citep{2008Wyatt,2013Panic,2015Wyatt}.  Around older stars the tenuous, dusty remnants of these protoplanetary discs remain observable through the presence of excess emission above that of the stellar photosphere at infrared wavelengths \citep{2001Habing,2003Decin,2007Wyatt,2011Kains}. These faint, gas-poor discs are replenished by the mutual collision of unseen planetesimals -- hence these systems are called ``debris discs'' \citep{1993Backman,2014Matthews}.  To date, the vast majority of known debris discs remain unresolved.  

Most of our current knowledge of the evolution of debris discs has been derived from the analysis of multi-wavelength photometry \citep[i.e. their spectral energy distributions, or SEDs; e.g. ][]{2007Wyatt,2011Kains} and mid-infrared spectroscopy \citep{2006Chen,2006aBeichman, 2014Chen, 2015Mittal}.  

At the most basic level, the observational properties of debris discs are limited to their temperature and brightness (fractional luminosity, $L_{\rm dust}/L_{\star}$), derived from fitting simple (modified) blackbody models to their SEDs \citep{2008Wyatt}. By definition, debris discs have fractional luminosities lower than 10$^{-2}$ \citep{2000Lagrange,2015Wyatt}.  

Drawing parallels to the Solar system, the debris discs are most commonly divided into warm and cool discs, analogous to the Solar system's Asteroid Belt \citep{2010Nesvorny} and Edgeworth-Kuiper Belt \citep{2012Vitense}, repsectively. Recent surveys of nearby stars reveal an incidence of cool dust around Sun-like stars (FGK spectral types) of 20 $\pm$ 2 per cent \citep{2013Eiroa,2016Mont}.  Observational constraints place the faintest known debris discs at fractional luminosities ($L_{\rm dust}/L_{\star}$) of a few $\times10^{-7}$ \citep{2013Eiroa}, around five to ten times brighter than the predicted brightness of the Edgeworth-Kuiper Belt \citep{2012Vitense}.  It is therefore worth noting that, with current technology, we would so far have failed to detect the Solar system's debris components; we might therefore assume that the Solar system is dust free.

The SED of an unresolved debris disc system can be used to broadly infer the structure of the underlying planetary system. Assuming thermal equilibrium between the dust and incident stellar radiation means that the radial location of dust derived from a disc's blackbody temperature is the minimum separation at which the dust could lie from the star. Studies of resolved debris discs revealed that the disc extent is often much greater than that predicted from the dust temperature \cite[e.g.][]{2011Marshall, 2012Wyatt, 2012RodZuck}. A trend between host star luminosity and dust temperature for A- to G-type stars has been noted \citep{2013Booth, 2013Morales} with larger discs, relative to the blackbody radius, observed around later type stars. Modelling of a sample of 34 extended debris discs parameterised this relationship, allowing more plausible estimates of the true extent of unresolved discs to be made \citep{2014Pawellek, 2015PawKri}. 

For many debris discs, modelling their excesses using a single component often fails to accurately reproduce the observed SEDs.  For this reason, a growing number of studies have modelled systems using two (or more) components. Such systems are often inferred to have multiple dust-producing planetesimal belts; however, the migration of dust and planetesimals within a system could also explain such observations \citep{2011Morales,2014Duchene,2014KennedyW}. A more detailed understanding of the architectures of these systems is limited by the fundamental degeneracies between the assumed disc radial location, minimum dust grain size and dust composition \citep[e.g.][]{2008Lohne,2010Krivov}.  Such knowledge is particularly important in placing circumstellar debris in context as a component of the planetary systems, alongside planets, around their host stars \citep{2012Wyatt,2012Maldonado,2015Maldonado,2014aMarshall,2015MM,2015WitMar}.

The \textit{Herschel} Space Observatory \citep{2010Pilbratt}, with its large, 3.5-m primary mirror, offered an unprecedented opportunity to spatially resolve circumstellar emission at far-infrared and sub-millimetre wavelengths. The \textit{Herschel} Photodetector Array Camera and Spectrometer \citep[PACS;][]{2010Poglitsch} had imaging capabilities in three wavebands centred on 70, 100 and 160~$\micron$.  

The analysis of resolved images of debris discs in the far-infrared, made possible by the high angular resolution of PACS (5\farcs8 FWHM at 70 $\micron$), in conjunction with more densely sampled source SEDs greatly refined models of nearby debris disc systems.  As a result, \textit{Herschel}/PACS led to the detection of many new debris disc systems. In addition, thanks to the excellent spatial resolution by \textit{Herschel}, it was able to resolve approximately half of the discs imaged by PACS \citep[e.g.][]{2010Liseau,2011Marshall,2012Lohne,2013Eiroa,2013Booth,2014Ertel,2014MatthewsHR8799,2014bMarshall}. These results made possible, for the first time, the quantification of the relationship between the presence of debris and the presence of planets \citep{2013Bryden}. It also allowed the architectures of a number of planetary systems to be more precisely determined, as a direct result of the measurement of the orientation of the debris in those systems \citep{2014Greaves,2014Kennedy}.

HD 105211 ($\eta$ Cru, HIP~59072) was identified as a debris disc host star through observations at 70 $\micron$ by the \textit{Spitzer} Space Telescope's \citep{2004Werner} Multiband Imaging Photometer for Spitzer instrument \citep[MIPS;][]{2004Rieke}. The detection image revealed a bright disc with extended emission \citep{2006bBeichman}. Here we analyse the higher angular resolution \textit{Herschel}/PACS images of this target in combination with available data from the literature, seeking to better determine the architecture of this nearby, bright debris disc.

In Section \ref{s:OA} we describe the observations obtained and the associated analysis for HD 105211.  In Section \ref{s:RD}, the results of the analysis of the SED, disc images and radial profiles at PACS 70/100/160 $\micron$ are presented along with measurements of the disc observational properties: temperature $T_{\rm dust}$, radial extent $R_{\rm dust}$ (both from the blackbody assumption and resolved/deconvolved images), and fractional luminosity $L_{\rm dust}/L_{\star}$. In Section \ref{s:D}, we discuss the state of the disc in comparison with other studied systems and the potential asymmetry present. In Section \ref{s:C}, we summarise our conclusions and discuss our plans for future work in studying this system. 

\section{Observations and analysis}\label{s:OA}

In this section, we present the observational data for the HD 105211 system.  This includes the characterisation of the host star with the fitting of a stellar photosphere model, a summary of the ancilliary photometry compiled for the SED, and a description of the \textit{Herschel} observations, their reduction and analysis.  The compiled SED for HD 105211, a scaled photospheric model, and a blackbody fit to the dust excess are presented in Figure \ref{fig:SED}.

\subsection{Stellar parameters}

HD 105211 ($\eta$ Crucis, HIP 50972) is a nearby \citep[d = 19.76 $\pm$ 0.05 pc,][]{2007vanL}, main-sequence star. Currently, HD 105211 is defined as both a spectroscopic binary \citep[The Washington Visual Double Star Catalog,][]{2001Mason} and a main sequence star with a spectral type of F2 V \citep{2006Gray} and it was originally classified as a yellow-white giant \citep[F2 III,][]{1971Jorg}. HD 105211 has been estimated to have an age between 1.30 to 3.99 Gyr with an effective temperature of 6950 K \citep{2006Gray} and a sub-solar metallicity [Fe/H] of -0.37.  A summary of the stellar physical properties are given in Table \ref{tab:StarParam}.

 \begin{table}
 \caption{Physical Properties of HD 105211.  \label{tab:StarParam}}
 \centering 
 \begin{tabular}{lll}
  \hline
  Parameter & Value & Ref.\\
  \hline\hline
  Distance [pc] & 19.76 $\pm$ 0.05 & 1\\
  Right Ascension [h:m:s] & $12:06:52.89$ & 1 \\
  Declination [d:m:s] & $-64:36:49.42$ & 1 \\
  Proper Motions (RA,Dec) [mas/yr] &  33.88 $\pm$ 0.12 & 1 \\ 
                                   & -37.02 $\pm$ 0.10 & 1 \\
  Spectral \& Luminosity Class & F2 V & 2 \\
  $V$, B.C. [mag] & 4.142 $\pm$ 0.004, -0.05 & 3\\
  $B$ - $V$ [mag] & 0.353 $\pm$ 0.007 & 3\\
  Bolometric Luminosity [L$_\odot$] & 6.6 $\pm$ 0.2 & 4\\
  Radius [$R_{\odot}$]& 1.66 $\pm$ 0.04 & 4\\
  Mass [$M_{\odot}$] & 1.63 $\pm$ 0.04 & 4 \\
  Temperature [K] & 7244 $\pm$ 50 & 4 \\
  Surface Gravity, $\log g$ [cm/s$^{2}$] & 4.21 $\pm$ 0.03 & 4 \\
  Metallicity [Fe/H] & -0.37 $\pm$ 0.18 & 5, 6, 9 \& 10 \\
  $v$ sin $i$ [km/s] & 46.1 $\pm$ 2.3 & 7 \\
  Age [Gyr] & 1.30-3.99 & 8, 9 \& 10\\
    \hline
 \end{tabular}

\medskip
\raggedright
 {References.} (1) \cite{2007vanL}; (2) \cite{2006Gray}; (3) \cite{1993Turon}; (4) \cite{1999AllPri}; (5) \cite{2006bBeichman}; (6) \citep{1995Marsakov}; (7) \cite{2012Ammler};  (8) \cite{2001Feltzing}; (9) \cite{2002Ibuk}; (10) \cite{2004Nord}.
 
\end{table}

The stellar photospheric contribution to the SED was modelled using an appropriate model taken from the Castelli-Kurcz atlas\footnote{Castelli-Kurcz models can be obtained from: http://www.stsci.edu/hst/observatory/crds/castelli\_kurucz\_atlas.html} \citep{2004CK}. The chosen photospheric model was scaled to the observations at optical and near-infrared wavelengths between 0.5 and 10 $\micron$, weighted by their uncertainties, using a least squares fit.

\subsection{Ancillary data}

\begin{table}
\caption{Photometry of HD 105211 used in SED modelling. \label{tab:Photometry}}
\centering
\begin{tabular}{lrr}
\hline
Wavelength  & Flux & Reference \\
  $[\micron]$ & [Jy] & \\
\hline\hline
0.349 & 24.020 $\pm$ 0.240 & 1 \\
0.411 & 62.400 $\pm$ 0.620 & 1 \\
0.440 & 69.910 $\pm$ 6.450 & 2 \\
0.466 & 72.700 $\pm$ 0.770 & 1 \\
0.546 & 69.910 $\pm$ 6.450 & 1 \\
0.550 & 79.990 $\pm$ 7.380 & 2 \\
0.64 & 92.200 $\pm$ 0.460 & 3 \\
0.79 & 88.400 $\pm$ 0.400 & 3 \\
1.26 & 59.160 $\pm$ 7.650 & 4 \\
1.60 & 45.850 $\pm$ 5.930 & 4 \\
2.22 & 29.790 $\pm$ 1.920 & 4 \\
3.40 & 14.840 $\pm$ 1.340 & 5 \\
8.28 &  2.870 $\pm$ 0.118 & 6 \\
9    &  2.270 $\pm$ 0.070 & 7 \\
12   &  1.420 $\pm$ 0.180 & 5 \\
13   &  1.105 $\pm$ 0.039 & 8 \\
18   &  0.690 $\pm$ 0.030 & 7 \\
22   &  0.434 $\pm$ 0.007 & 5 \\
24   &  0.368 $\pm$ 0.025 & 9 \\
27   &  0.296 $\pm$ 0.015 & 10 \\
31   &  0.228 $\pm$ 0.011 & 8 \\
33   &  0.222 $\pm$ 0.022 & 10 \\
35   &  0.214 $\pm$ 0.038 & 10 \\
70   &  0.559 $\pm$ 0.073 & 9 \\
70   &  0.733 $\pm$ 0.063 & 11 \\
100  &  0.728 $\pm$ 0.096 & 11 \\
160  &  0.564 $\pm$ 0.095 & 11 \\
\hline
\end{tabular}

\medskip
\raggedright
\textbf{References.} (1) \cite{1998HauckMerm}; (2) \textit{Hipparcos} catalogue, \cite{1997ESA}; (3) \cite{1978MorelMagnanet}; (4) \cite{1985Epchtein}; (5) \textit{WISE} all-sky survey, \cite{2010Wright}; (6) AKARI IRC all-sky survey, \cite{2010Ishihara}; (7) \textit{MSX} catalogue \citep{2003Egan}; (8) \textit{Spitzer}/IRS, \citep{2015Mittal}; (9) \textit{Spitzer}/MIPS, \cite{2014Chen}; (10) \textit{Spitzer}/IRS, this work; (11) \textit{Herschel}/PACS, this work. 
\newline
\end{table}

For the purposes of modelling HD 105211's SED, the \textit{Herschel} PACS photometry were supplemented with a broad range of observations from the literature spanning optical to far-infrared wavelengths.  A summary of the compiled photometry is given in Table \ref{tab:Photometry}.

The optical Johnson \textit{BV} photometry were taken from the \textit{Hipparcos} \& Tycho Catalogues \citep{1997ESA}, whilst the near-infrared \textit{RI} and \textit{JHK} photometry were taken from the catalogues of Morel \& Magnenat \citep{1978MorelMagnanet} and Epchtein \citep{1985Epchtein}, respectively. These measurements were supplemented by Str\"omgren $ubvy$ photometry from \cite{1998HauckMerm}.

Mid-infrared photometry included in the SED were taken from the \textit{Midcourse Space eXperiment} \citep[\textit{MSX};][]{2003Egan}, AKARI IRC all-sky survey at 9 and 18 $\micron$ \citep{2010Ishihara}, the \textit{WISE} survey at 3.4, 12, and 22 $\micron$ \citep[the 4.6 $\micron$ band measurement was saturated][]{2010Wright}, and a \textit{Spitzer} MIPS measurement at 24 $\micron$ \citep{2014Chen}. Colour corrections were applied to the AKARI IRC measurements assuming a blackbody temperature of 7000 K (factors of 1.184 at 9 $\micron$ and 0.990 at 18 $\micron$), and also applied to the \textit{WISE} data assuming a Rayleigh-Jeans slope (factors of 1.0088 at 12 $\micron$ and 1.0013 at 22 $\micron$).

A \textit{Spitzer} InfraRed Spectrograph \citep[IRS;][]{2004Houck} low-resolution spectrum spanning $\sim$7.5 to 38 $\micron$ (PID 2324, PI Beichman, AOR 16605440) was taken from the CASSIS\footnote{The Cornell Atlas of Spitzer/IRS Sources (CASSIS) is a product of the Infrared Science Center at Cornell University, supported by NASA and JPL.} archive \citep{2011Lebouteiller}. The IRS spectrum was scaled to match the model photosphere at wavelengths $<$ 10 $\micron$, where no excess emission from the disc was expected, by a least-squares fitting process. The rescaling factor for the IRS spectrum was 0.94, in line with other works \citep[e.g.][]{2009Chen, 2014bMarshall}. For inclusion in the SED modelling process, the IRS spectrum was binned with a weighted average mean (with associated uncertainty) at 27, 30, 33 and 36 $\micron$ in order to trace the rise in the excess emission from the dust above the photosphere at mid-infrared wavelengths (see section \ref{DSED} for details). 

\subsection{\textit{Herschel} data}

At far-infrared wavelengths, the previous \textit{Spitzer}/MIPS 70 $\micron$ measurement \citep{2006bBeichman} was supplemented by the new \textit{Herschel}/PACS observations. HD 105211 was observed as part of the Herschel Open Time 1 programme ot1\_sdodson\_1 (PI: S. Dodson-Robinson; \textit{`A study into the conditions for giant planet formation'}), the results of which were presented in \cite{2016DodRob}. The log of \textit{Herschel} observations is given in Table \ref{tab:obsers}. 

\begin{table}
 \caption{Summary of \textit{Herschel}/PACS observations of HD 105211. \label{tab:obsers}}
 \centering
 \begin{tabular}{llll}
  \hline
 Observation ID & OD & Wavelengths [$\micron$] & Duration [s]\\
  \hline\hline
1342262373/74 & 1355 & 70/160 & 465\\
1342262371/72 & 1355 & 100/160 & 445 \\
  \hline
 \end{tabular}
 \newline
\end{table}

PACS scan map observations of HD 105211 were taken in both 70/160 and 100/160 channel combinations.  A summary of the PACS observations is presented in Table \ref{tab:obsers}.  Each scan map was conducted at a slew rate of $20\arcsec$ per second (medium scan speed) and is composed of eight scan legs of $3\arcmin$ long, each separated by $4\arcsec$. The target was observed twice, at instrument orientation angles of 70\degr and 110\degr, for each channel combination. This resulted in a total of two scan maps each for 70 and 100 $\micron$ and a total of four scan maps for 160 $\micron$. 

Reduction of the \textit{Herschel}/PACS data was carried out in the Herschel Interactive Processing Environment\footnote{{\sc hipe} is a joint development by the Herschel Science Ground Segment Consortium, consisting of ESA, the NASA Herschel Science Center, and the HIFI, PACS and SPIRE consortia.\\ http://www.cosmos.esa.int/web/herschel/hipe-download}  \cite[{\sc hipe};][]{2010Ott} user release 13.0.0 and PACS calibration version 69 (the latest available public release at the time), using the standard reduction script starting from the level 0 products.  All available images (i.e. two each at 70 and 100 $\micron$, and four at 160 $\micron$) were combined to produce a final mosaic for each waveband. The image scales for the mosaiced images were 1\farcs2 per pixel for the 70 and 100 $\micron$ images, and 2\farcs4 per pixel for the 160 $\micron$ image.  To remove large scale background emission, the images were high-pass filtered with widths of 15 (70/100 $\micron$) and 25 (160 $\micron$) frames, corresponding to spatial scales of $62\arcsec$ and $102\arcsec$, respectively. 

Fluxes were measured using an aperture photometry task implemented in {\sc{Matlab}}. In the 70 and 100 $\micron$ images we adopted aperture radii of $24\arcsec$, in order to encompass the whole disc within the measurement aperture. At 160 $\micron$ the disc is contaminated by the presence of bright, nearby background structure to the west. We therefore use a smaller aperture radius of $17\arcsec$ at 160 $\micron$, despite the larger beam size (11\farcs8 FWHM at 160 $\micron$ compared to 5\farcs8 FWHM at 70 $\micron$), to reduce the contribution of this background to the measurement. These values were scaled by appropriate aperture correction factors of 0.877, 0.867 and 0.758 at 70, 100 and 160 $\micron$ photometry, respectively \citep{2014Balog}. We note that whilst applying aperture corrections derived for point sources may be contentious when dealing with extended sources, this method has been widely used in the literature \citep[e.g.][]{2010Liseau,2011Marshall,2013Booth,2013Roberge,2014Ertel,2016Marshall}.

The \textit{Herschel} flux measurements in each waveband were also colour corrected, treating the stellar and dust contributions separately. Appropriate factors were taken from the PACS Photometer colour correction release note\footnote{http://herschel.esac.esa.int/twiki/pub/Public/PacsCalibrationWeb/ cc\_report\_v1.pdf} corresponding to their temperature values of 6,950 K for the star (interpolating between tabulate values) and 50 K for the dust. At 70, 100, and 160 $\micron$ we used values of 1.016 (0.982), 1.033 (0.985) and 1.074 (1.010) for the stellar (dust) contributions, respectively. The final fluxes used in the modelling were further corrected for losses (1 to 2 per cent) induced by the high-pass filtering process, as detailed in \cite{2012Popesso}. The PACS fluxes and associated uncertainties are shown in Table \ref{tab:Photometry}.

The sky background in each mosaic was estimated from the median values of ten randomly placed 10$\times$10 pixel sub-regions of each mosaic. These regions were masked to avoid HD 105211, any bright background objects (e.g. CL Cru -- see Section \ref{A:CL_Cru}), and the borders of the mosaics (where noise increases due to lower coverage). The median value was scaled to the revelant aperture size corresponding to the target radius to yield the sky background contribution, which was then subtracted from the measured flux (before correction, as detailed above). The sky noise was estimated from the standard deviation of the ten sub-regions. 

\begin{figure}
\includegraphics[width=\columnwidth]{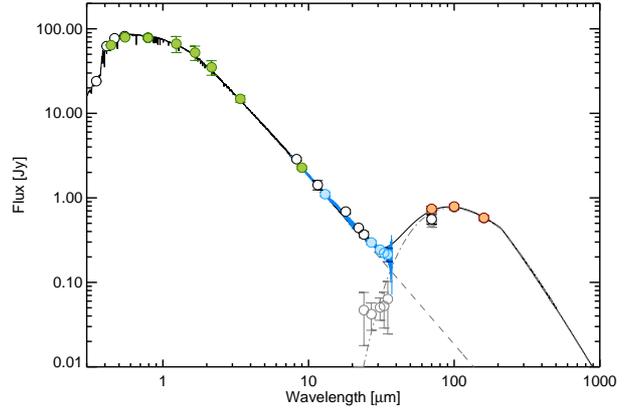}
 \caption{SED of HD 105211. Green data points are the optical and near-infrared fluxes used to scale the stellar photosphere model. Blue data points are fluxes calculated from binning the \textit{Spitzer} IRS spectrum. Orange data points are the \textit{Herschel} PACS observations. White data points are ancilliary data from the literature. Grey data points are excess (total - star) fluxes at wavelengths $\geq 20 \micron$. The dashed grey line is the stellar photosphere, the dot-dash grey line is the dust contribution, and the solid black line is the combined star+disc model. The dust contribution, estimated as a 50 K blackbody, has been modified by a factor of $(\lambda/210)^{-1}$ at wavelengths beyond 210 $\micron$, although the absence of sub-millimetre data for the disc leaves the fall-off of the SED unconstrained. \label{fig:SED}}
\end{figure}

\section{Results}\label{s:RD}

Here we present the results of our analysis of HD 105211's disc. We examine the PACS images for evidence of extended emission, before applying a deconvolution routine to better determine the disc extent and geometry. The additional photometric points are combined with ancillary data and the disc architecture measured from the images to model the disc, fitting the excess emission with both modified blackbody and power law disc models from which we deduce the dust grain properties.

\subsection{Images}\label{RD:DiscImages}

\begin{figure*}
\centering
\includegraphics[width=2\columnwidth]{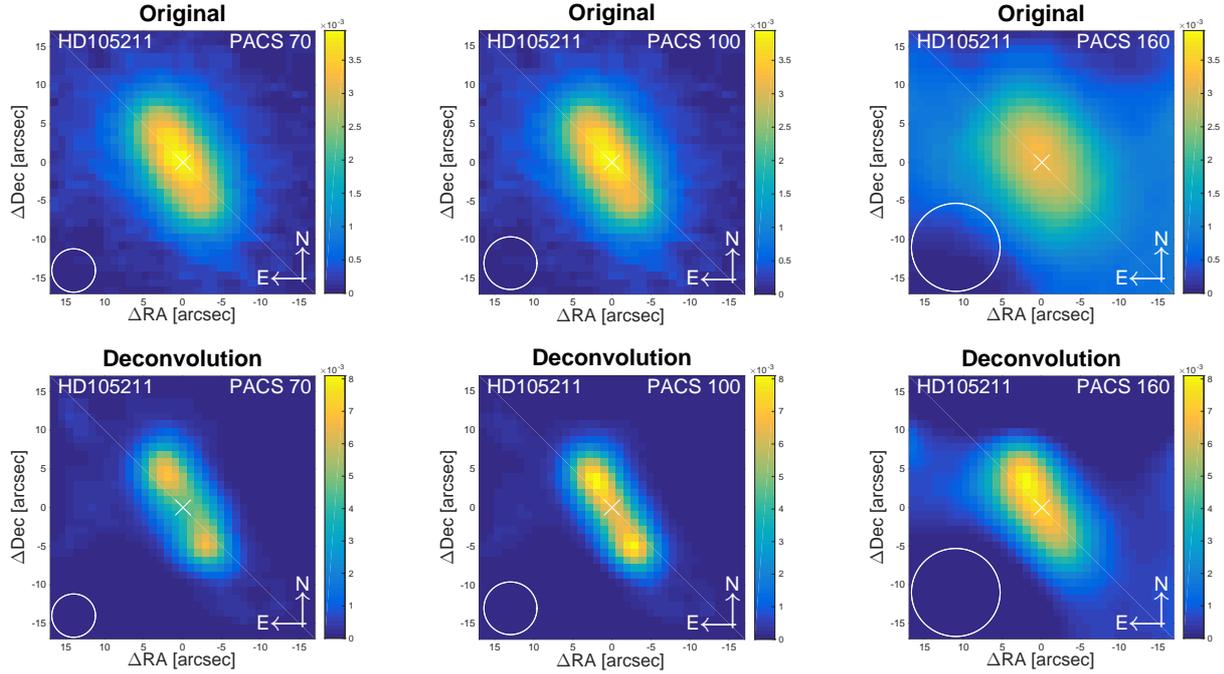}
 \caption{Original Resolved Images (Top Row), Lucy-Richardson deconvolution images on original images with star subtraction (Bottom Row). The star's position is marked with a white cross. Instrument FWHM beam is represented by the circle in the bottom left hand corner. Image orientation is north up, east left. The images have the same scale of $1\arcsec$ pixel$^{-1}$.  The colour scale bar is in units of Jy/arcsec$^{2}$. The image r.m.s. (in Jy arcsec$^{-2}$) values are $1.0\times10^{-4}$ at 70 $\micron$, $1.4\times10^{-4}$ at 100 $\micron$, and $3.0\times10^{-4}$ at 160 $\micron$. \label{fig:stamps}}
\end{figure*}

HD 105211 was resolved by the \textit{Herschel}/PACS instrument \citep{2010Poglitsch} showing the extent of the disc for wavelengths 70, 100 and 160 $\micron$ (see top row in Figure \ref{fig:stamps}). The 70 and 100 $\micron$ maps were originally 1 pixel to 1\farcs2, whilst the 160 $\micron$ map was 1 pixel to 2\farcs4.  All maps were scaled to 1 pixel to $1\arcsec$ for subsequent analysis.    

\subsubsection{Stellar position}

The optical position of HD 105211, at the epoch of the \textit{Herschel} observations, is $12^h6^m52\fs81 \ {-64}\degr36\arcmin49\farcs91$; utilising the proper motions from the re-reduction of \textit{Hipparcos} data \citep{2007vanL}. Two methods were used to estimate the star's position in the maps: finding the position of the peak pixel brightness in all maps and calculating the centre of a fitted 2D Gaussian profile, allowing for rotation and ellipticity, of the 70 and 100 $\micron$ maps. Fitting a 2D Gaussian to the 160 $\micron$ PACS map was not feasible due to background contamination surrounding HD 105211. The peak pixel and the centre of the disc profile are both within $\sim$~$2\farcs5$ from the optical position, which is just outside the \textit{Herschel} absolute pointing accuracy of $2\arcsec$ at the 1-$\sigma$ level \citep{2014ESan}.  In the deconvolution presented here, we have assumed that the star's position for the 70 and 100 $\micron$ maps is the corresponding centre of 2D Gaussian disc profile fit.
 
\subsubsection{Radial profiles and deconvolution}
\label{RD:radprof}
In the original maps, the extent of the semi-major and -minor axes were derived from the FWHM of model 2D Gaussian profiles along the major and minor axes.  In the 70 and 100 $\micron$ maps the extents were found to be similar, with values of 160.4 $\pm$ 1.7 au and 91.0 $\pm$ 0.8 au, respectively.  However, the 160 $\micron$ map yielded a larger estimate of the disc extent along both axes of 192 $\pm$ 32 au and 141 $\pm$ 14 au. The disc inclination angle was determined from the inverse cosine of the ratio of the semi-minor axis to the semi-major axis. The inclination was consistent in the 70 and 100 $\micron$ maps at 55.4 $\pm$ $5.2 \degr$. The inclination angle for 160 $\micron$ map, 42.7 $\pm$ $8.3 \degr$,  is just within the uncertainty limits when compared with the values derived in the shorter wavelength maps.  This could be attributed to both the rising level of background contamination to the west of the source, and the larger PACS beam FWHM of 11\farcs8 at 160~$\micron$ (cf 5\farcs8 at 70 $\micron$), comparable to the disc extent, such that the disc is poorly resolved in the 160 $\micron$ image.

A reference point spread function (PSF) was used to estimate the stellar contribution to the images, and to deconvolve the original maps. Here, we have adopted a similar technique to that used in \cite{2010Liseau} and \cite{2011Marshall}.  Observations of the chosen reference PSF, $\alpha$ B{\"o}otis (HD 124897, Arcturus) were reduced in the same way as the HD~105211 images and rotated to the same telescope pointing angle.  The stellar photosphere contribution was modelled separately by scaling the PSF to the flux estimate corresponding to 70, 100 and 160 $\micron$ wavelengths of the photosphere model respectively.  The stellar contribution was centred on the previously determined stellar position in each image and subtracted from the three original maps.  After stellar subtraction the Lucy-Richardson method \citep{1972Richardson,1974Lucy} was then used to deconvolve the original maps with the corresponding instrument PSFs (see bottom row in Figure \ref{fig:stamps}).

\begin{figure}
\includegraphics[width=1.05\columnwidth]{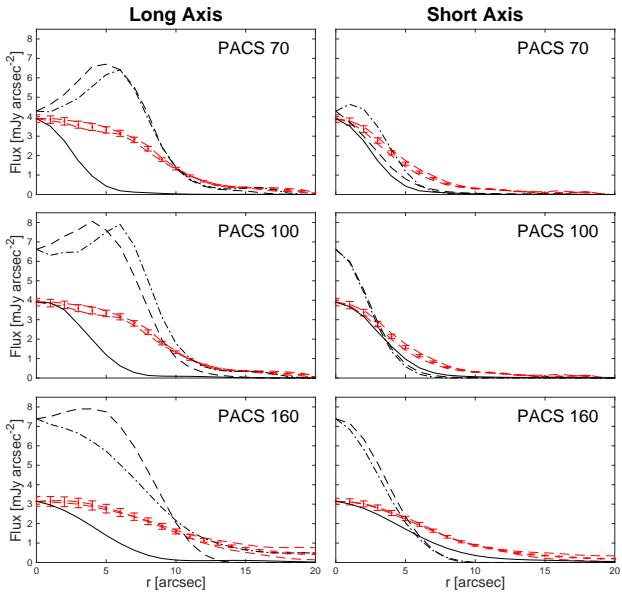}
 \caption{Radial Profiles for HD 105211 along the semi-major (left) and semi-minor (right) axes of the disc. The red dashed lines are the measured profile either side of the disc centre along each axis; the solid red errors bars mark the mean position at each point with the associated uncertainty.  The black dashed-dotted lines and the black dashed lines are the deconvolved disc profiles  along the SW arm and NE arm respectively for the long axis; along the NW extent and SE extent respectively for the short axis.  The solid black line is the subtracted stellar profile that has been scaled to the peak emission of the original image for reference.}
 \label{fig:radial}
\end{figure}

After deconvolution, the semi-major and -minor axes of the disc were determined by fitting an ellipse to the region of the map, centred on HD 105211, which exceeded a 3-$\sigma$ threshold. These measured values are larger than the original PACS images due to the difference between the 2D Gaussian and ellipse fits.  The semi-major and -minor axes were used to estimate the inclination angle, as before.  The inclination angles measured in the deconvolved images are consistent with values of $70.1 \pm 2.3 \degr$ at 70 $\micron$ and $70.7 \pm 2.2 \degr$ at 100 $\micron$. The increased size of the inclination angle from the original PACS maps can be attributed to the deconvolution resolving the axis of the disc more precisely. However, the inclination angle for 160 $\micron$ of $60.1 \pm 7.7 \degr$ is not consistent due to the poorer resolution of the original image. Radial profiles are shown in Figure \ref{fig:radial} and a summary of associated measurements is shown in Table \ref{tab:rp}.

 \begin{table}
 \caption{Measurements of HD 105211's disc extent and inclination. \label{tab:rp} }
 \centering
 \begin{tabular}{llll}
  \hline
 & 70$\micron$  &   100$\micron$     & 160$\micron$  \\
  \hline
  \hline
Original PACS Maps & & &\\
Semi-major [\arcsec] & 8.11 $\pm$ 0.81 & 8.12 $\pm$ 0.81 & 9.74 $\pm$ 1.62\\
Semi-major [au] & 160.3 $\pm$ 16.0 & 160.5 $\pm$ 16.0 & 192 $\pm$ 32 \\
Semi-minor [\arcsec] & 4.61 $\pm$ 0.35 & 4.60 $\pm$ 0.35 & 7.14 $\pm$ 0.69\\
Semi-minor [au] & 91.1 $\pm$ 6.9 & 90.9 $\pm$ 6.9 & 141 $\pm$ 14 \\
Inclination Angle [\degr] & 55.4 $\pm$ 5.2 & 55.5 $\pm$ 5.1  & 42.7 $\pm$ 8.3 \\
Position Angle [\degr] & 30.1 $\pm$ 0.5 & 30.1 $\pm$ 0.5 & --\\
\hline
Deconvolved Images & & & \\
Semi-major [\arcsec] & 9.92 $\pm$ 0.50 & 9.93 $\pm$ 0.50 &  9.82 $\pm$ 0.98 \\
Semi-major [au] & 196.0 $\pm$ 9.9 &  196.2 $\pm$ 9.9 & 194.0 $\pm$ 19.4\\
Semi-minor [\arcsec] & 3.38 $\pm$ 0.17 & 3.29 $\pm$  0.16 &  4.90 $\pm$ 0.49 \\
Semi-minor [au] & 66.8 $\pm$ 3.4 & 65.0 $\pm$  3.2 & 96.8 $\pm$ 9.7 \\
Inclination Angle [\degr] & 70.1 $\pm$ 2.3 & 70.7 $\pm$ 2.2  & 60.1 $\pm$ 7.7 \\
Peak-to-Peak [\arcsec] & 10.5 $\pm$ 0.5 & 9.9 $\pm$ 0.5 & -- \\
Peak-to-Peak [au] & 207.5 $\pm$ 9.8 & 195.6 $\pm$ 9.8 & -- \\
NE Arm [\arcsec]  & 4.70 $\pm$ 0.2  & 4.0 $\pm$ 0.2 & 3.4 $\pm$ 0.3 \\
NE Arm [au] & 93 $\pm$ 4.0 & 79 $\pm$ 4.0 &  67 $\pm$ 6.0 \\
SW Arm [\arcsec]  & 5.8 $\pm$ 0.3 & 5.9 $\pm$ 0.3 & -- \\
SW Arm [au] & 115 $\pm$ 5.8 & 117 $\pm$ 5.8 &  --  \\
  \hline
 \end{tabular}
 \newline
 \end{table}

\subsubsection{Asymmetry}

The deconvolved images at 70 and 100 $\micron$ reveal distinct peaks along the NE and SW arms. The disc extent along the major-axis of the 70 and 100 $\micron$ deconvolutions were measured from peak-to-peak. The peak-to-peak extent was not observed in the 160 $\micron$ deconvolution, as there is only a peak emission along the NE arm. The NE and SW arms were measured from the star centre position to peak emission of the disc.  The NW and SE extents are the half of the FWHMs of the deconvolved disc profile from star centre position.  Examining the 70 and 100 $\micron$ deconvolutions, the NE clump is larger and closer to the star's centre position (86.0 $\pm$ 5.7 au) in contrast to its SW counterpart (116.0 $\pm$ 8.2 au) but both comparable in brightness, clearly showing an asymmetry in the disc. However, the disc at 160 $\micron$ has no peak emission on either side of disc centre but still shows asymmetry, with a smoothly decreasing brightness profile from the NE arm to the SW arm. A summary of disc profile measurements is shown in Table \ref{tab:rp}.

\subsection{Spectral energy distribution}\label{DSED}

\subsubsection{Modified blackbody}

The SED reveals no significant emission at mid-infrared wavelengths $\leq 20~\micron$, as shown in Fig. \ref{fig:SED}. We therefore determine the disc observational properties of temperature, (blackbody) radial extent, and fractional luminosity ($T_{\rm dust}$, $R_{\rm dust}$, $L_{\rm dust}/L_{\star}$) through fitting a single component, modified blackbody model to the photometry at wavelengths where a significant ($>$ 3-$\sigma$) excess was measured. Due to the absence of any photometry at wavelengths beyond 160 $\micron$, the break wavelength ($\lambda_{\rm 0}$) and sub-millimetre slope ($\beta$) of the modified blackbody model are unconstrained and for the purposes of plotting the best-fit model we assume a break wavelength of 210 $\micron$, and a $\beta$ of 1, as typical values for this purpose \citep{2008Wyatt}. 

A least-squares fit to the photometry weighted by the uncertainties produces a fractional luminosity of $L_{\rm dust}/L_{\star}\,=\,7.4 \pm 0.4 \times10^{-5}$, in line with previous estimates of the disc brightness \citep{2006bBeichman}, and a best-fit temperature of $T_{\rm dust}\,=\,49.7 \pm 0.6$ K. The temperature is equivalent to a blackbody radius of $R_{\rm dust}\,=\,80.2 \pm 2.4$ au. A comparison with the resolved extent from the images (80 or 120 au) reveals that the ratio of the actual to blackbody radii, $\Gamma\,=\,$ 1 to 1.5. This is surprisingly low for a star of 6.6 $L_{\odot}$, as we expect a value of $\Gamma$ in the region 2.65$^{+0.57}_{-0.24}$ by following the models of \cite{2015PawKri}. The radiation blow-out radius for dust grains \citep{1979BurnsLamySoter} around HD~105211 is 3.5~$\mu$m. The small value of $\Gamma$ leads us to infer that the disc emission is therefore dominated by large, cool dust grains and that any population of smaller grains, that would be warmer at a given stellocentric distance \citep{2008Krivov}, is thus low. The HD 105211 system has not been imaged in scattered light; however, aperture polarisation observations in the optical (SDSS g$^{\prime}$ and r$^{\prime}$ measurements) have been made and the analysis of this data is ongoing (Cotton D. priv. comm).

Compact residual emission centred on the star is visible in the 70 $\micron$ disc-subtracted image (see top right panel of Figure \ref{fig:pl}).  This emission is weak, constituting a 2-$\sigma$ detection at 70 $\micron$.  However, it could represent the marginal detection of a warm inner belt to the HD~105211 system.  We therefore estimate the properties of this inner belt based on the available mid-infrared photometry between 24 and 37 $\micron$, along the with the 70 $\micron$ measurement. Using an error-weighted least squares fit to these measurements, we obtain a temperature of 215 $\pm$ 43 K (equivalent to radial distance of 4.3 $\pm$ 1.9 au, assuming the same properties as the cold disc) with a fractional luminosity of approximately $5 \times 10^{-6}$.

\subsubsection{Radiative transfer}

Combining the SED and resolved imaging, we now apply a more physically complex model to the available data. Here we simultaneously fit the extended emission and SED with a radiative transfer model using a power law size distribution of dust grains assuming a dust composition of astronomical silicate \citep{2003Draine}. The debris disc is assumed to lie at some distance from the star in an annulus between $R_{\rm in}$ and $R_{\rm in} + \Delta R$, where $R_{\rm in}$ was constrained to lie between 50 and 150~au, and $\Delta R$ was a free parameter. The exponent of the surface density of the disc $\alpha$ was allowed to vary between 2.0 and -5.0. As noted in the previous section there was some evidence for an asymmetry in the disc from the deconvolved images; here we assumed an axisymmetric model in the first instance. The constituent grains are represented by a power law size distribution between $a_{\rm min}$, a free parameter, and $a_{\rm max}$, which is fixed at 1 mm, with an exponent $q$. Due to the lack of sub-millimetre photometry, we assumed the value of $q$ lies in the range 3 to 4, typical of the theoretically expected \citep{1969Dohnanyi,2005PanSari,2012Gaspar,2012PanSchlicht} and observed \citep{2015Ricci,2016Macgregor} values for debris disc systems. 

The brightness profiles of the disc along its major and minor axes for all three PACS images were used to represent the resolved emission in the fitting process. The profiles were measured as described in Section \ref{RD:radprof}. For the weighting of the fit, each image was given equal weighting to the SED in determining the best-fit (i.e. the SED and each image each contribute 25 per cent of the best-fit).

We found a simultaneous best fit model to the images and SED with a reduced $\chi^{2}$ of 1.61 (six free parameters). The inner edge of the disc lies at $R_{\rm in} = 87.0^{+2.5}_{-2.3}~$au, close to the blackbody radius inferred for the disc. The disc was found to be broad, $\Delta R = 100^{+10}_{-20}~$au, with an outwardly decreasing radial surface density exponent $\alpha = -1.00^{+0.22}_{-0.24}$. The minimum dust grain size was calculated to be $5.16^{+0.36}_{-0.35}~\micron$, mostly dictated by the fit to the rising \textit{Spitzer} IRS spectrum; this is perhaps surprising given how close the disc inner edge lies to the blackbody radius. The exponent of the size distribution $q = 3.90^{+0.10}_{-0.14}$, lies within the assumed range, and is consistent with theoretical expectations assuming a velocity dependent dispersion of collision fragments \citep{2012PanSchlicht}. A dust mass of $2.45^{+0.05}_{-0.13} \times10^{-2}M_{\oplus}$ is inferred from the model, but the validity of this model-derived value is highly uncertain in the absence of sub-millimetre photometry. The results of the modelling are summarised in Table \ref{tab:pl}.

\begin{table}
 \caption{Power law fit. \label{tab:pl} }
 \centering
 \begin{tabular}{llll}
  \hline
Parameter &  Range  &   Distribution   &  Fit  \\
  \hline\hline
$R_{\rm in}$ (au) & 50 -- 150 & Linear & $87.0^{+2.5}_{-2.3}$ \\
$\Delta R$ (au) & 10 -- 100 & Linear & 100$^{+10}_{-20}$ \\
$\alpha$ & 2.0 -- -5.0 & Linear & $-1.00^{+0.22}_{-0.24}$ \\
$a_{\rm min}$ ($\micron$) & 1.0 -- 30.0 & Logarithmic & $5.16^{+0.36}_{-0.35}$ \\
$a_{\rm max}$ ($\micron$) & Fixed & n/a & 1000 \\
$q$ & 3.0 -- 4.0 & Linear & $3.90^{+0.10}_{-0.14}$ \\
Composition & astron. sil. & Fixed & --- \\
$M_{\rm dust}$ ($\times10^{-4}M_{\oplus}$) & Free & n/a & $2.45^{+0.05}_{-0.13}$ \\
$\chi^{2}_{\rm red}$ & Free & n/a & 1.61 \\
  \hline
 \end{tabular}
 \newline
 \end{table}

\begin{figure*}
\centering
\subfigure{\includegraphics[width=0.50\textwidth,trim={0cm 0cm 0cm 0cm},clip]{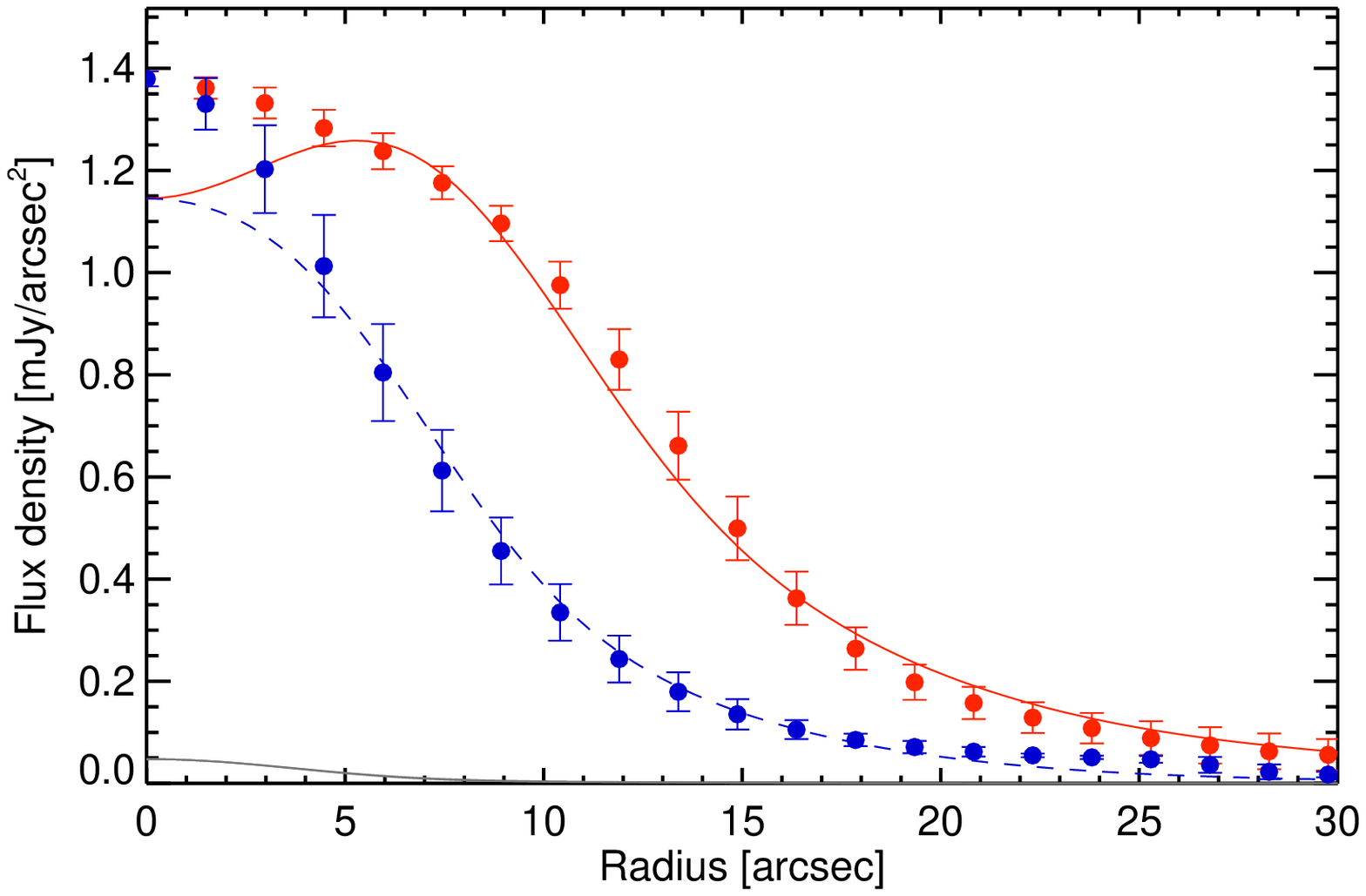}}
\subfigure{\includegraphics[width=0.42\textwidth,trim={5cm 0cm 0cm 0cm},clip]{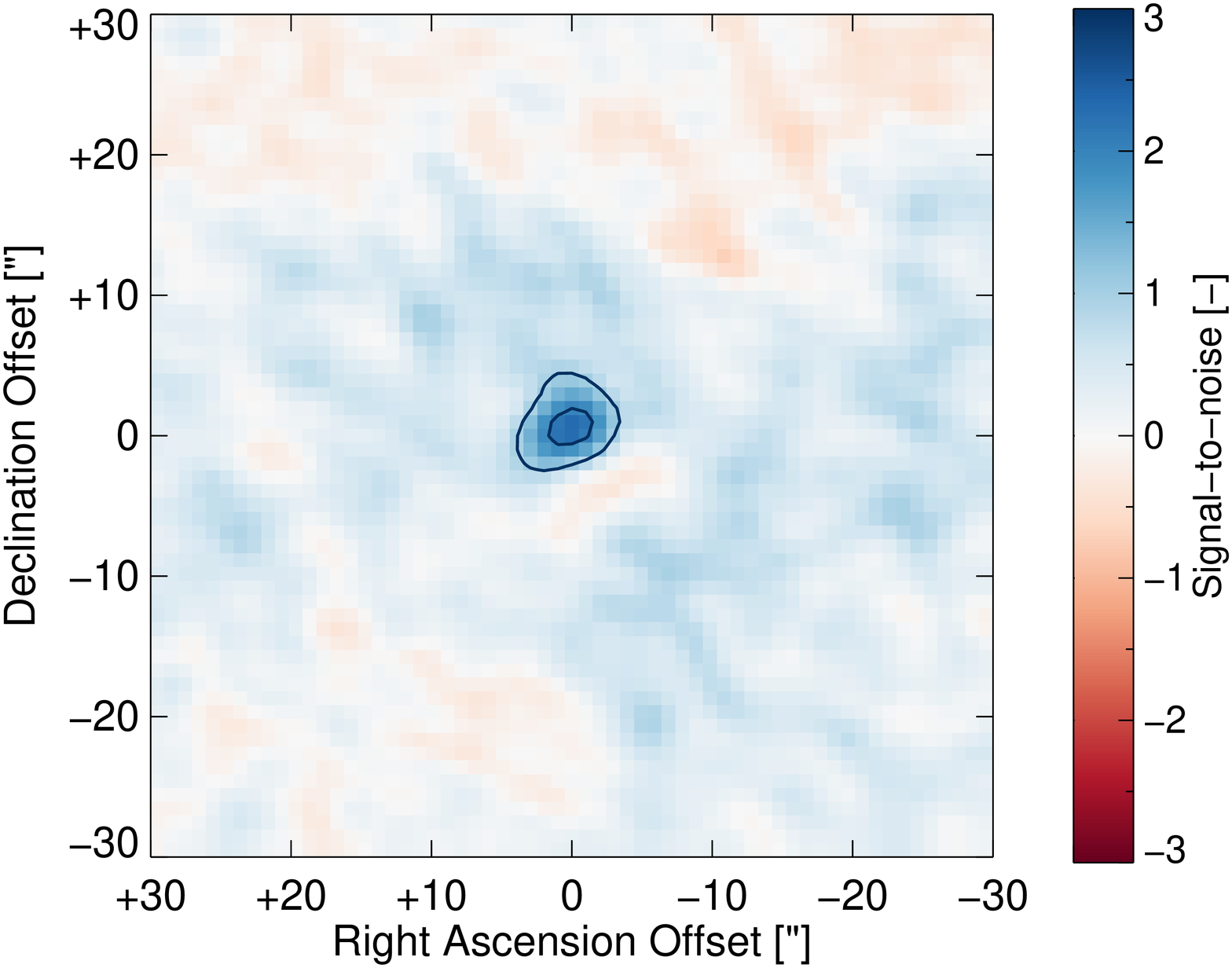}}\\
\vspace{-0.5in}
\subfigure{\includegraphics[width=0.50\textwidth,trim={0cm 0cm 0cm 0cm},clip]{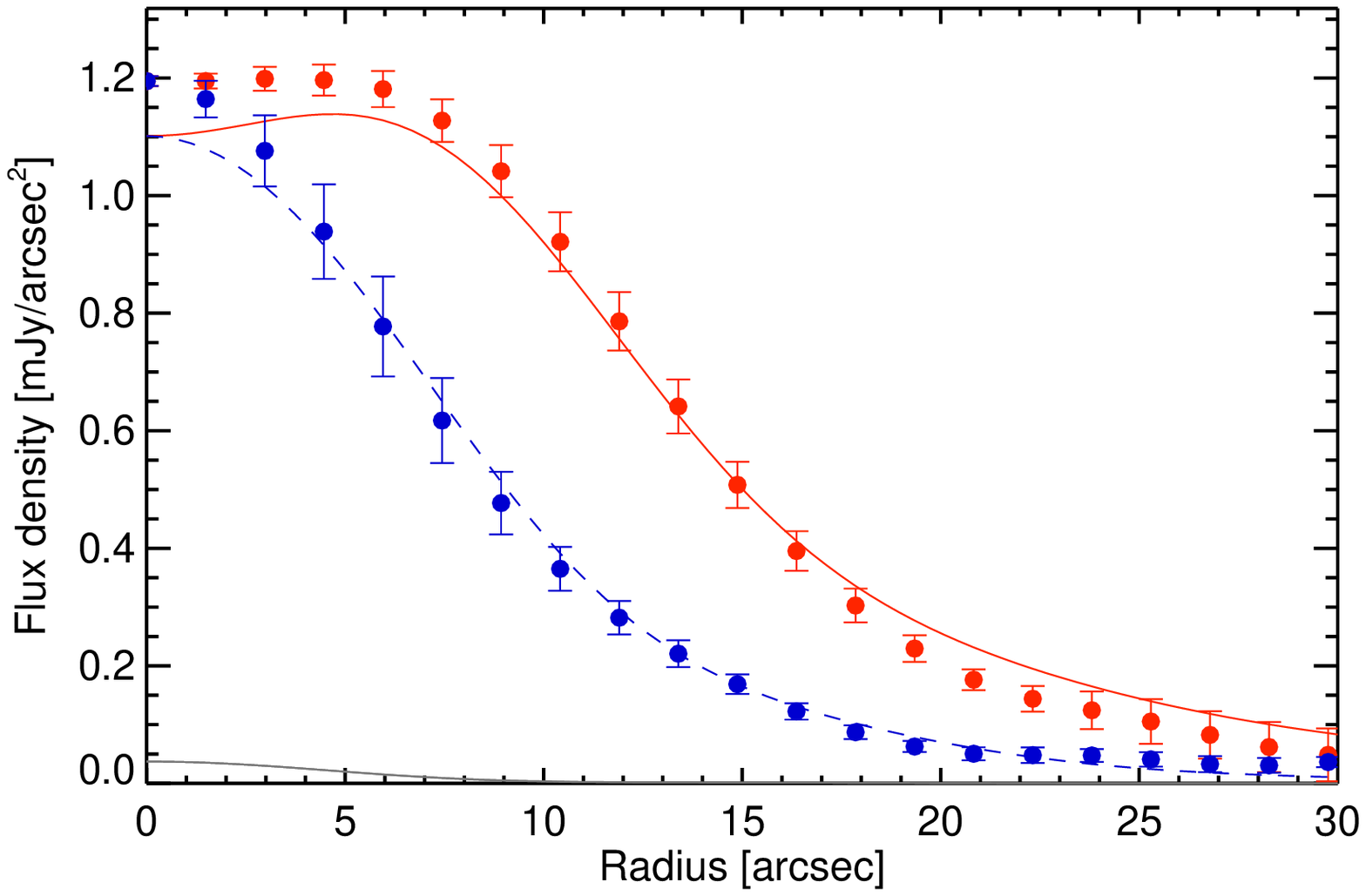}}
\subfigure{\includegraphics[width=0.42\textwidth,trim={5cm 0cm 0cm 0cm},clip]{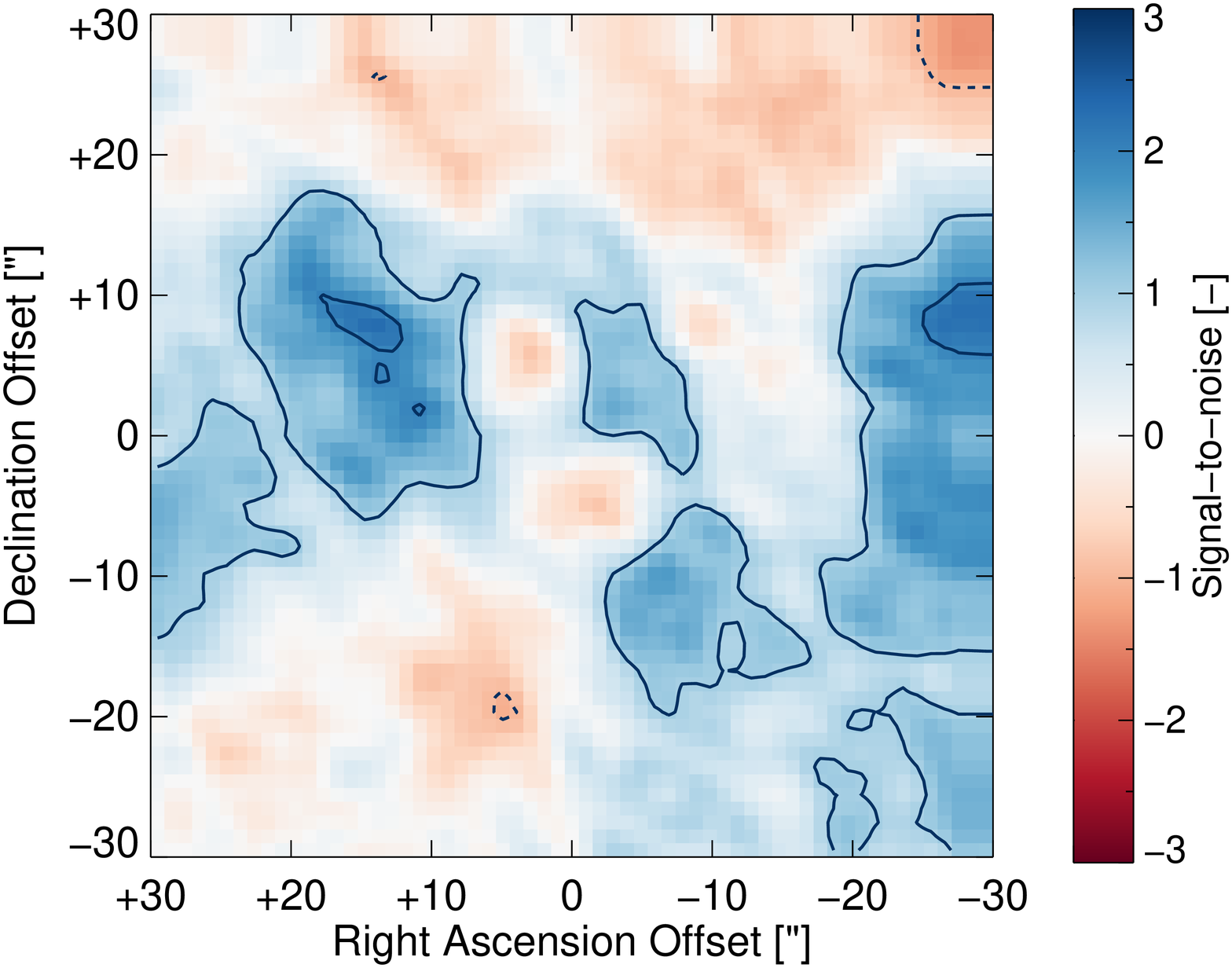}}\\
\vspace{-0.5in}
\subfigure{\includegraphics[width=0.50\textwidth,trim={0cm 0cm 0cm 0cm},clip]{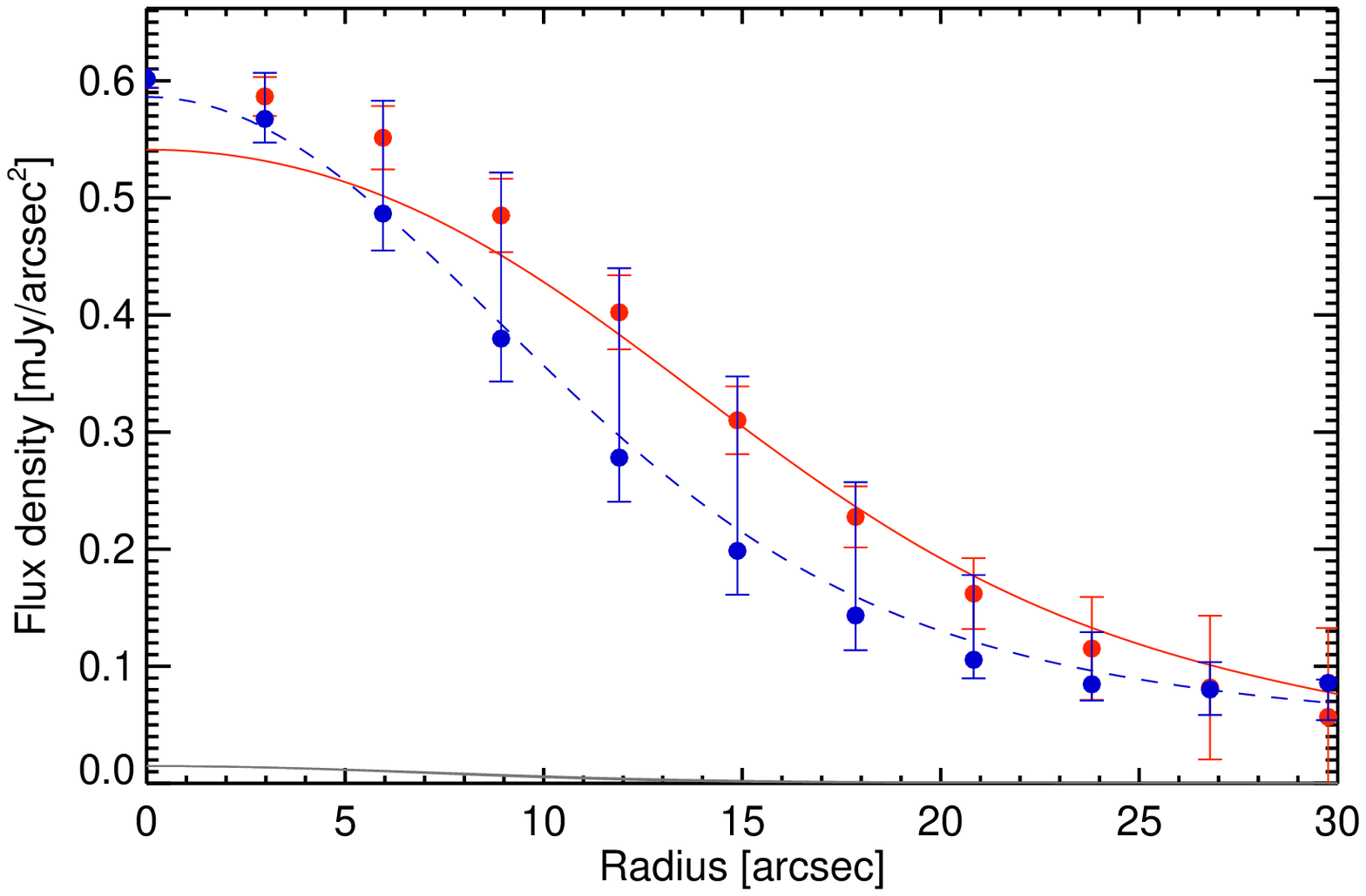}}
\subfigure{\includegraphics[width=0.42\textwidth,trim={5cm 0cm 0cm 0cm},clip]{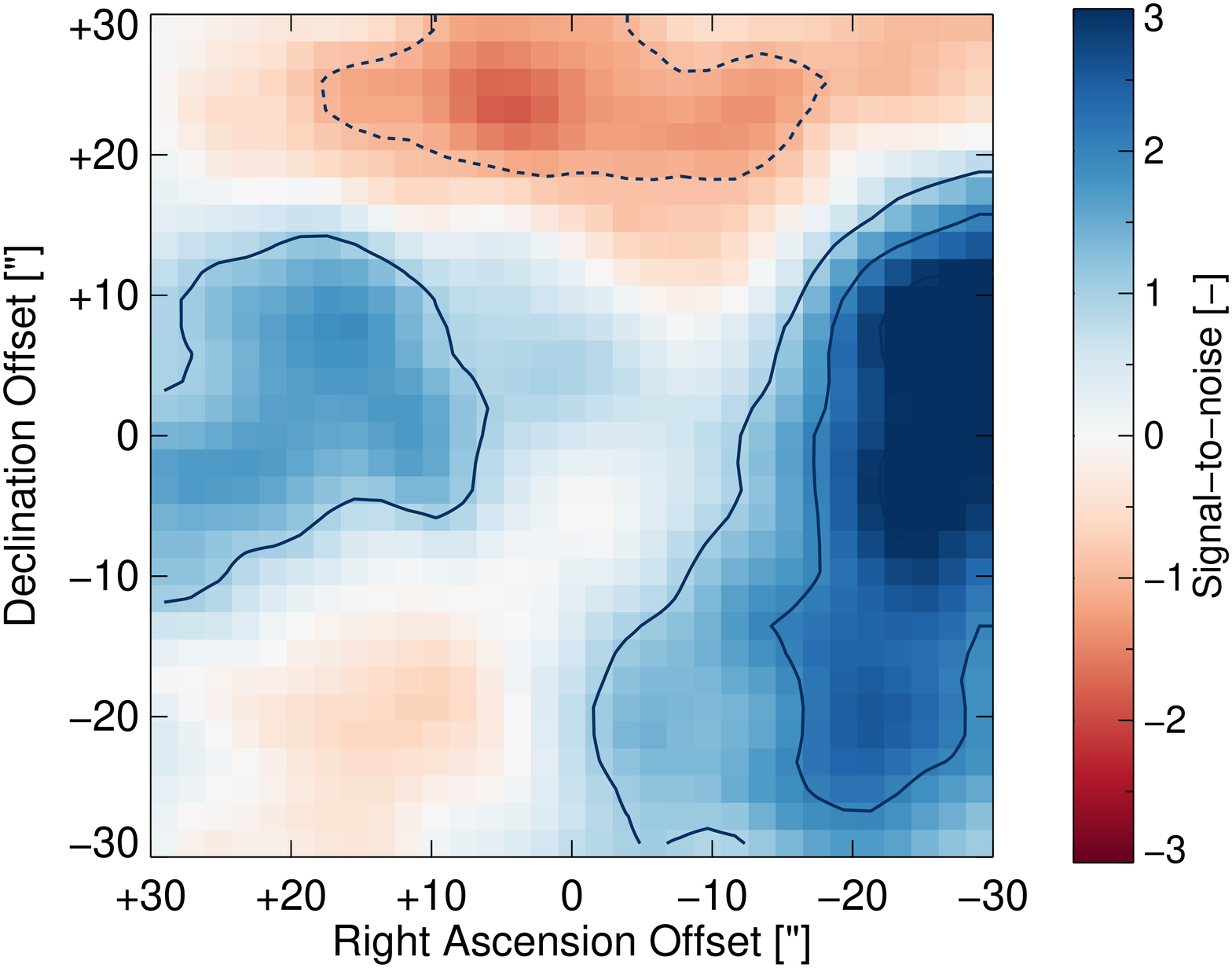}}\\
\caption{Results of fitting the disc extended emission at 70, 100, and 160~$\mu$m (top to bottom). \textit{Left}: Brightness profiles along the major and minor axes of HD 105211's disc.  Data points are the average of the east and west extent of the disc at the given separation. The red solid line denotes the major axis of the model, the blue dashed line denotes its minor axis, and the grey line is the stellar contribution. \textit{Right}: Residual signal-to-noise maps after models of the disc architecture convolved with the appropriate PSF are subtracted from the observed images. Red areas denote negative flux, while blue areas denote positive flux. Contours are $\pm$1-, $\pm$2-, and $\pm$3-$\sigma$, with dashed lines denoting negative values. Orientation is north up, east left. The image scale is $1\arcsec$ per pixel for 70 and 100 $\micron$, and $2\arcsec$ per pixel for 160 $\micron$. \label{fig:pl}}
\end{figure*}

In Figure \ref{fig:pl}, we present the \textit{residual} emission after subtraction of the disc model from the observations. Overall the model replicates the observations well, with no significant residual flux at the disc position in any of the three images. This is interesting given the potential asymmetry inferred from the deconvolved images. In this instance, the evidence lies with an axisymmetric disc capable of replicating both the structure and thermal emission of the disc, such that we cannot confirm the presence of asymmetry in the disc with the data to hand.

However, at 70 \micron, a 2-$\sigma$ peak is present at the stellar position. The model brightness profile has a deficit of flux compared to the observations close to the stellar position (see top left panel of Figure \ref{fig:pl}), this could be used to infer the presence of an additional, unresolved warm component to the disc. A two component model was previously used to fit the \textit{Spitzer} IRS spectrum in \cite{2015Mittal}, but the SED shows no significant remaining excess emission that would require the presence of an additional disc component (see Figure \ref{fig:SED}), so we opted not to include one in our analysis. Radial drift of material from the outer belt toward the star could account for the presence of this faint emission.

\subsection{CL Cru}\label{A:CL_Cru}

\begin{table}
\caption{Photometry of CL Cru. \label{tab:CLCruPhotometry}}
\centering
\begin{tabular}{lc}
 \hline
Wavelength  & Flux \\
  $[\micron]$ & [Jy] \\
\hline\hline
70   &  0.558 $\pm$ 0.040 \\
100  &  0.297 $\pm$ 0.023 \\
160  &  0.155 $\pm$ 0.024 \\
 \hline
 \end{tabular} \\
\end{table}

The Mira variable star CL Cru (IRAS 12043-6417) is clearly visible in the 70 and 100 $\micron$ PACS mosaics, lying in the northwest corner, but is barely detected in the 160 $\micron$ mosaic. Note, this object is not visible in in Figure \ref{fig:stamps} as it lies outside the cropped region presented there. Its flux measurements are included here as they may be of interest to those involved in research on evolved stars. Circular apertures of 4, 5, and $8\arcsec$ radius were used to measure the fluxes at 70, 100 and 160 $\micron$, respectively. The background contribution and uncertainty was estimated from a annulus spanning 15 to $25\arcsec$ from the source. CL Cru has no evidence of extended emission, based on a 2D Gaussian fit to the source, and we therefore extract the fluxes with radii for optimal signal-to-noise. The measurements were aperture corrected, with correction factors of 0.487, 0.521 and 0.527 \citep{2014Balog} for the 70, 100 and 160 $\micron$ measurements, respectively. At 70 and 100 $\micron$ the noise is dominated by the calibration uncertainty, whereas at 160 $\micron$ the sky noise is the dominant contribution. The aperture corrected (but not colour corrected) flux density measurements are given in Table \ref{tab:CLCruPhotometry}.

\section{Discussion}\label{s:D}

Here we place the results of our analysis into context, and examine the origins and implications for a potential asymmetry of the disc around HD 105211. 

\subsection{State of the disc}

The radiative transfer model determined that the minimum grain size for the disc (assuming a composition of pure astronomical silicate) was $5.16^{+0.36}_{-0.35}~\micron$. This result is only marginally larger than the blowout size for grains in the disc. This is unexpected due to the extent of the disc determined from the resolved images and SED fitting being $87.0^{+2.5}_{-2.3}$~au, close to the blackbody radius of 80 au, such that we would perhaps expect a larger minimum grain size to account for such a low ratio in measured to blackbody radii. 

Additionally, the ratio of observed radius to the blackbody radius of the disc ($\Gamma = 1.1$) is smaller than the expected ratio of $\sim$~2.5 when correlated with stellar luminosity \citep{2015PawKri}. HD 105211's debris disc has been previously noted as being bright in thermal emission for its age \citep[$> 1$~Gyr,][]{2011Kains} when compared to the expected trends for Sun-like stars \citep[e.g.][]{2001Spangler,2003Decin}. We therefore have a sketch of the disc around HD 105211 as being both bright (albeit not anomalously so) with large dust grains similar in size found in HD 207129 \citep{2011Marshall,2012Lohne}. The inner edge of the disc is close to the minimum (blackbody) extent with a broad disc component of about 100 au.  The inferred disc extent (down to the blackbody radius) requires a minimum grain size above the blow-out radius for the star's luminosity. The large dust grains will likely be broken down into smaller, warmer fragments that could persist around HD~105211.  These fragments would then migrate inward by radiation transport processes, which could account for the marginal excess seen in the 70 $\micron$ image.

\subsection{Potential asymmetry}

The deconvolved images revealed an asymmetry in the separation of the two ansae from the stellar position, but the structure and emission from the disc around HD~105211 can be effecitvely replicated with a single, axisymmetric annulus. We do not therefore propose that the disc is asymmetric, but here investigate the possibilities of such an asymmetry, and discuss future directions to refine our understanding of the disc architecture.  

Asymmetries are often cited as the result of perturbation by an unseen planetary companion influencing the dynamics of the dust-producing planetesimals \citep[e.g.][]{2009Chiang,2014Faramaz,2016ThilMad}. The disc asymmetry is observed to be brighter at longer wavelengths on the side closer to the star, which suggests the dust there is colder, perhaps due to trapping of larger grains by a perturbing body. Additional observations at higher angular resolution in the far-infrared to fill in the SED between the \textit{Spitzer} and \textit{Herschel} photometry would be invaluable in pinning down the location of the star relative to the disc, and determining if a second warm component is present in the inner regions of the disc.

There are other possible scenarios to explain the presence of an asymmetry. Given HD~105211's location, lying close to the galactic plane ($b \sim -2\degr$), background contamination along the line of sight could offer an explanation. Bright, extended emission is present across the images, particularly at 160 $\micron$, but there is little at either 70 or 100 $\micron$ where the asymmetry is most obvious. Additionally, a background point source would need to have a flux $\sim$ 50 mJy, lying at a separation of $\leq 9\arcsec$ (along the semi-major axis) from the star to replicate the observed asymmetry. Considering extragalactic contamination \citep{2013Sibthorpe}, the binomial probability of such an occurence is low ($P = 0.009$) and should therefore give us confidence in the real presence of the asymmetry. However, HD 105211 lies close to the galactic plane and such estimates, based on galaxy counts, should be considered a lower limit in such a scenario.

Alternatively, interaction between the disc and interstellar medium might also induce an asymmetry, as has been seen for a number of systems resolved in scattered light \citep[e.g. HD~15115, HD~61005; ][]{2012Rodigas,2009Maness,2014Schneider}. 

Obtaining more observations at sub-millimetre wavelengths of HD~105211 would enable the size distribution of grains and their mass to be precisely determined; and these derived values used in combination would confirm the disc's asymmetry.  This would also clarify if it is exclusive to smaller grains that are susceptible to radiation pressure and stellar winds or if it persists for larger grains that more accurately trace the parent planetesimal belts of the disc.

\subsection{Comparison with Dodson-Robinson et al.}

Our analysis of HD~105211 is generally consistent with the results of \cite{2016DodRob}. However, we differ in several of the particulars. We note that our single component model, while broad, matches the detected excesses measured from \textit{Spitzer} IRS through \textit{Herschel} wavebands, as shown in Figure \ref{fig:pl}. Most importantly, we find no evidence for significant warm excess emission from the system at 24~$\mu$m. This is likely due to the different photometry used to scale the stellar photosphere models in each work; as \cite{2016DodRob} used 2MASS $JHK_{s}$ (which are of poor quality) and AllWISE data (which suffered saturation not present in the all-sky survey). The lack of significant warm excess in our analysis led us to adopt a single component disc model, leading to the result that we require a broad disc to replicate both the SED and radial profiles. \cite{2016DodRob}, with the freedom of an additional dust component, found the outer disc to be more narrowly confined at a larger radial distance (175 $\pm$ 20~au). The validity of both models can be tested by observations at high spatial resolution by ALMA (which will resolve the width of the outer cool component) and JWST (searching for a distinct warm component). 

\cite{2016DodRob} also determined a smaller flux density for HD~105211 at 160~$\mu$m (440~$\pm$~146~mJy) than in this work. There is significant contamination from bright extended structure within the image at 160~$\mu$m, so the determination of the correct flux is fraught with difficulty. We note that our values are consistent within uncertainties and that obtaining longer wavelength observations, either from SOFIA/HAWC+ at 210~$\mu$m, or ALMA in the sub-millimetre would better determine the shape of the SED and confirm whether or not the disc emission had deviated from a blackbody at 160~$\mu$m.

\section{Conclusion}\label{s:C}

HD 105211 is nearby star (d = 19.76 $\pm$ 0.05 pc) and slightly more massive and more luminous than the Sun. It was reported to exhibit significant infrared excess by \textit{Spitzer}. In this work, we have presented new \textit{Herschel}/PACS far-infrared observations of the debris disc around HD 105211 at 70, 100 and 160 $\micron$. These images revealed extended emission from the disc along both its semi-major and semi-minor axes in all three wavebands, from which we determine the disc to have an inclination of 71 $\pm$ $2 \degr$ and a position angle of 30 $\pm$ $1 \degr$. 

Of particular interest is the presence of an intriguing asymmetry between the two sides of the disc after deconvolution, interpreted as ansae lying at 86 $\pm$ 5.7 au and 116 $\pm$ 8.2 au, from which we infer a disc eccentricity of 0.20 $\pm$ 0.03.  However, the asymmetry could not be considered significant when we modelled the structure and thermal emission of the disc and found that the disc was broad, 100 $\pm$ 20 au, starting at an inner edge of 87 $\pm$ 2.5 au. The star's location close to the galactic plane makes contamination a possibility, requiring further observation at higher angular resolution and longer wavelengths to better determine the architecture of this disc. 

Combining the disc SED with the multi-wavelength resolved images we fitted a power law disc model and derived a minimum grain size of $5.16^{+0.36}_{-0.35}~\micron$, assuming the disc lies in a single annulus around the star. The lack of any constraint on the sub-millimetre emission from the disc leaves the grain size distribution unconstrained in our model, again pointing the way to further observations to determine the dust properties for this system. 

\section*{Acknowledgements}
The authors wish to thank the anonymous referee for their feedback, which improved the clarity of this paper. They would also like to thank S. Dodson-Robinson for the use of her Herschel observing programme that included HD 105211. This research has made use of the SIMBAD database, operated at CDS, Strasbourg, France.  This research has made use of the VizieR catalogue access tool, CDS, Strasbourg, France.  The research has made use of NASA's Astrophysics Data System. JPM is supported by a UNSW Vice-Chancellor's Postdoctoral Fellowship. 




\bibliographystyle{mnras}
\bibliography{papers} 


\bsp	

\label{lastpage}
\end{document}